\documentclass[12pt,letter]{emulateapj}

\usepackage{apjfonts}
\usepackage{amsmath,amssymb}
\usepackage{xspace}
\usepackage{graphicx}
\usepackage{epstopdf}
\usepackage{graphicx}
\usepackage{epsfig}

\newcommand\bzk{\emph{BzK}}
\newcommand\simgt{\lower.5ex\hbox{\gtsima}}
\newcommand\simlt{\lower.5ex\hbox{\ltsima}}

\slugcomment{To appear in ApJ Letters}

\shorttitle{Iron emitting quasars at $z=2.5-3$}
\shortauthors{Feruglio et al. }

\begin{document}


\title{Discovery of strong Iron K$\alpha$ emitting Compton thick quasars at $z=2.5$ and 2.9}


\author{C. Feruglio\altaffilmark{1}, 
E. Daddi\altaffilmark{1}, 
F. Fiore\altaffilmark{2}, 
D. M. Alexander\altaffilmark{3}, 
E. Piconcelli\altaffilmark{2},
C. Malacaria\altaffilmark{4}
}

\altaffiltext{1}{Laboratoire AIM, CEA/DSM - CNRS - Universit\'e Paris Diderot,
       Irfu/Service d'Astrophysique, CEA Saclay, Orme des Merisiers,  91191 Gif-sur-Yvette Cedex, France}
\altaffiltext{2}{INAF - Osservatorio Astronomico di Roma, via Frascati 33, 00040 Monteporzio Catone, Italy}
\altaffiltext{3}{Department of Physics, Durham University, South Road, DH1 3LE, UK}
\altaffiltext{4}{Universita' La Sapienza, Piazzale Aldo Moro 5, 00185 Roma, Italy}
\email{chiara.feruglio@cea.fr}



\begin{abstract}
We report the detection of the 6.4 keV Iron K$\alpha$ emission line in two infrared-luminous, massive, star-forming 
\bzk~ galaxies  at $z=2.578$ and $z=2.90$ in the CDF-S.
The \emph{Chandra} 4 Ms spectra of \bzk4892 and \bzk8608 show a reflection dominated continuum with strong Iron lines, with rest-frame
equivalent widths EW$\sim$2.3 keV and 1.2 keV, respectively, demonstrating Compton thick obscuration of the central 
AGN. For \bzk8608 the line identification closely matches the existing photometric redshift derived from the stellar
emission. 
We use the observed luminosities of the Iron K$\alpha$ line, of the rest-frame mid-IR continuum and of the UV rest-frame narrow emission lines to infer intrinsic $L_{\rm 2-10~keV}\gtrsim 10^{44}$~erg~s$^{-1}$, about 1.0--2.5 dex larger than the observed
ones, hence confirming the presence of an absorber with $N_{\rm H} > 10^{24}$~cm$^{-2}$. The two \bzk~
galaxies have stellar masses of
$5\times10^{10}$~M$_\odot$ and, based on VLA 1.4~GHz and submm 870$\mu$m observations, they
appear to host vigorous starburst activity with $SFR\sim 300$-700~M$_\odot$~yr$^{-1}$
that is also optically thick. We estimate that the AGN might also conceivably 
account for an important fraction of the bolometric far-IR emission of the galaxies.
The implied volume density of Compton thick (CT) AGN with $L_{\rm 2-10~keV}>10^{44}$~erg~s$^{-1}$ is in agreement with predictions from X-ray background synthesis models.
These sources provide one of the first clearcut observations of the long-sought phase of simultaneous,
heavily obscured quasar and star formation activity, predicted by models of massive galaxy evolution at high redshifts.
\end{abstract}

\keywords{galaxies: general --- galaxies : active}

\section{Introduction}
Highly obscured, deeply embedded AGN might contribute significantly to the total accretion power in the Universe 
(Marconi et al. 2004) and are required to account for the spectrum of the X-ray background (XRB, Gilli et al. 2007). 
In particular, AGN obscured in the X-ray band by column densities larger than N$_H=10^{24}$ cm$^{-2}$
(Compton thick AGN) represent 20--25\%  of the AGN detected by Integral and SWIFT/BAT in the local universe
(Malizia et al. 2009; Burlon et al. 2010).

The identification of  highly obscured AGN is particularly challenging since, because of the high column densities,
the vast majority of their X-ray emission below rest frame 10~keV is absorbed, so that they 
are largely missed even in the deepest hard X-ray surveys available today.
At high redshift the quest for such objects is mainly pursued by indirect evidence, by selecting galaxies 
with high ratios of mid-infrared to optical, 
or mid-infrared to X-ray fluxes, and by deriving average X-ray properties by stacking techniques (e.g., Daddi et al.
2007b;
D07 hereafter;  Fiore et al. 2008, 2009; Alexander et al. 2005; 2008; Treister et al. 2010, Georgantopoulos et al. 2009, Eckart et al. 2010, Donley et al. 2010). These studies have
suggested that a substantial fraction of massive galaxies at $z>1$ host highly obscured, intrinsically luminous
AGN. 

Only a few tens of secure Compton thick AGN are known so far in the local Universe 
(Comastri et a. 2004, Dalla Ceca et al. 2008 and
references therein), and only for a handful of them 
a quasar-like intrinsic hard X-ray luminosity
has been inferred 
(L$_{2-10~keV}> 10^{44}$erg~s$^{-1}$; Braito et al. 2004, Piconcelli et al. 2010).
Their X-ray spectra are characterized by the presence of a strong Iron K$\alpha$ 6.4 keV fluorescent emission line,
with large equivalent width EW~$\gtrsim1$~keV, on a flat reflection-dominated continuum (Matt et al. 2000).
Not much evidence currently exists in the distant Universe
of AGNs showing prominent,  high EW Iron K$\alpha$ line: an IRAS selected
hyperluminous galaxy at $z=0.93$ (Iwasawa et al. 2005) and two possible detections in the CDFS
reported by Norman et al. (2002; a type 2 QSO at $z=3.7$) and by Tozzi et al. (2006; a galaxy at $z=1.53$);
both confirmed by Comastri et al. (2011).

In the following we present the significant 
\emph{Chandra} detection of the Iron K$\alpha$ emission line in two massive \bzk\ galaxies at
redshift z$=2.578$ and z$=2.90\pm0.10$ in GOODS-South, using the recently acquired 4~Ms dataset. 
We used for this purpose the entire set of 52 observations for a total exposure time of 4 Ms, available in the Chandra Data Archive (http://cxc.harvard.edu/cda/Contrib/CDFS.html).
We adopt a $\Lambda$CDM cosmology ($H_0=70$ km s$^{-1}$ Mpc$^{-1}$; $\Omega_M$=0.3; $\Omega_{\Lambda}=0.7$) 
and a Chabrier stellar initial mass function (IMF).

\begin{figure*}[t]
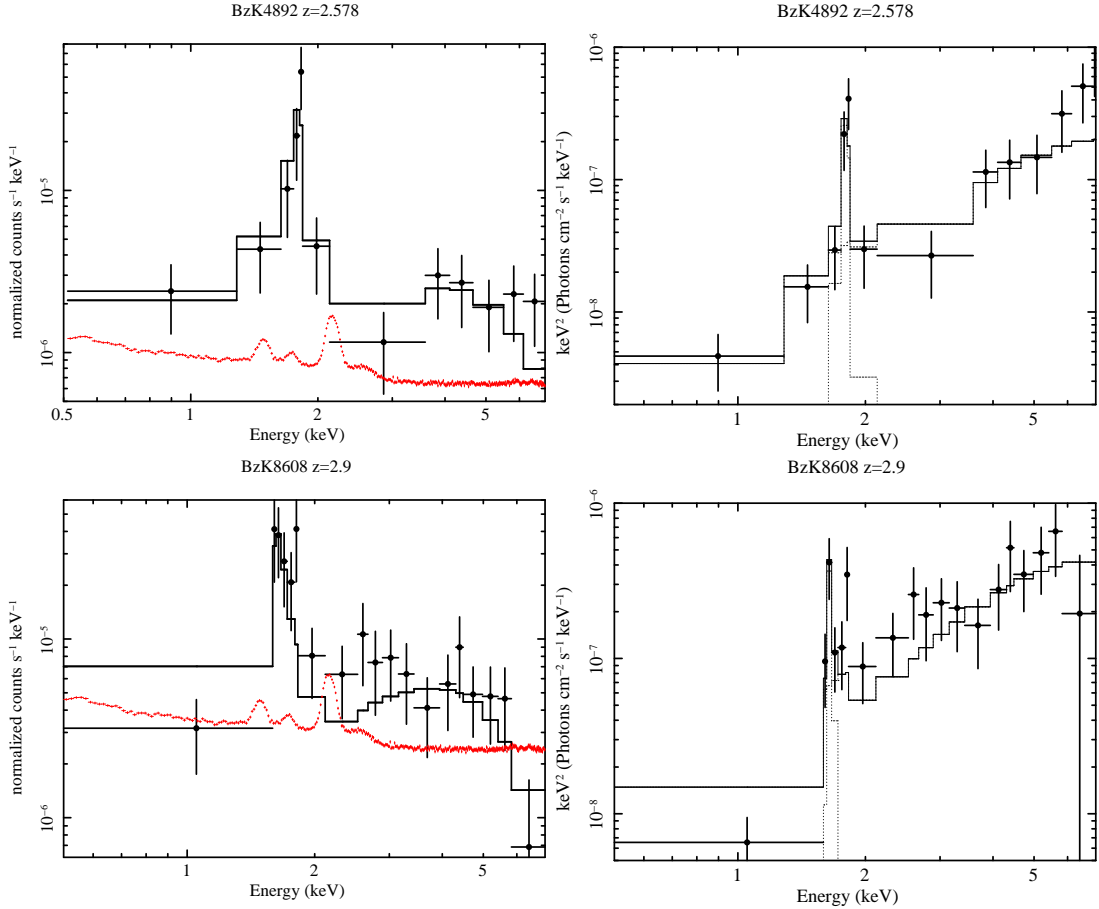

\centering
\includegraphics[angle=-90,scale=.3]{4892_spe_bgd_15.eps}
\includegraphics[angle=-90,scale=.3]{4892_euff.eps}\\
\includegraphics[angle=-90,scale=.3]{8608_spe_bgd.eps}
\includegraphics[angle=-90,scale=.3]{8608_eeuff.eps}
\caption{The spectra of \bzk4892 (upper panels) and \bzk8608 (lower panels), fitted with a pure reflection model 
plus an Iron K$\alpha$ 6.4~keV line. The spectra are shown both in observed counts units (left panels) and in
physical units (keV$^2$; right panels). Each bin corresponds to a 2$\sigma$ measurement. The red dotted line represents the average background extracted from the whole field, excluding sources (Fiore et al. 2011, in preparation).}
\end{figure*}

\begin{figure*}
\centering
\includegraphics[scale=0.7]{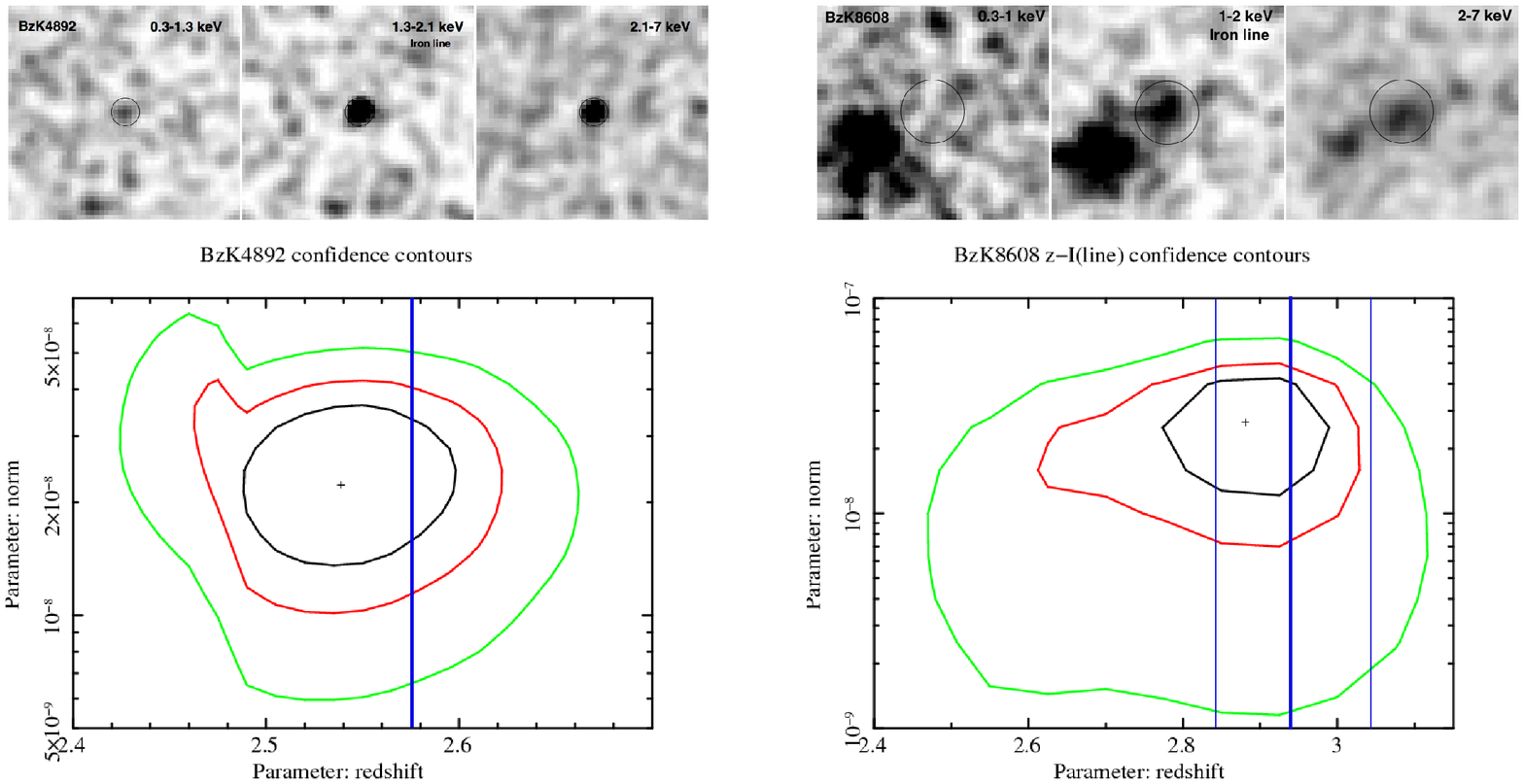}
\caption{Upper panels: \emph{Chandra} images of \bzk4892 and \bzk8608 in three energy bands. The middle images are
centered at the energy of the Iron K line. The circles show the extraction regions. Lower panels:
confidence levels (1, 2 and 3 $\sigma$) obtained from the fit of the Iron line. The vertical blue lines show the
optical spectroscopic/photometric redshift constraints.}
\label{fig:Xzp}
\end{figure*}

\section{X-ray photometric and spectral analysis}

We have been investigating the properties of \bzk-selected (Daddi et al. 2004), massive galaxies at redshift $1<z<3$ with faint or
no X-ray emission, with the aim of constraining their obscured AGN activity. We have focused in particular on 
objects showing also a power-law emission in the near-IR rest frame, an independent indication
for the presence of a luminous, obscured AGN. A large statistical sample of 750 X-ray undetected galaxies has been
assembled in the COSMOS field, and their X-ray properties investigated through stacking. As a control sample, 
we also studied in detail a smaller sample of 26 similarly selected
galaxies in GOODS-South, with  less than 200 net
counts in the 0.5-7 keV band in the CDFS 4 Ms data that would remain thus undetected in the shallower 
COSMOS \emph{Chandra} observations.
While the general results of these studies will be reported elsewhere (Feruglio et al. 2011, in preparation), we concentrate
here on two remarkable sources in our GOODS-S control sample,
for which the available \emph{Chandra} data allow the unambiguous detection of the Iron K$\alpha$ emission. 

\begin{figure*}
\centering
\includegraphics[angle=0,scale=.65]{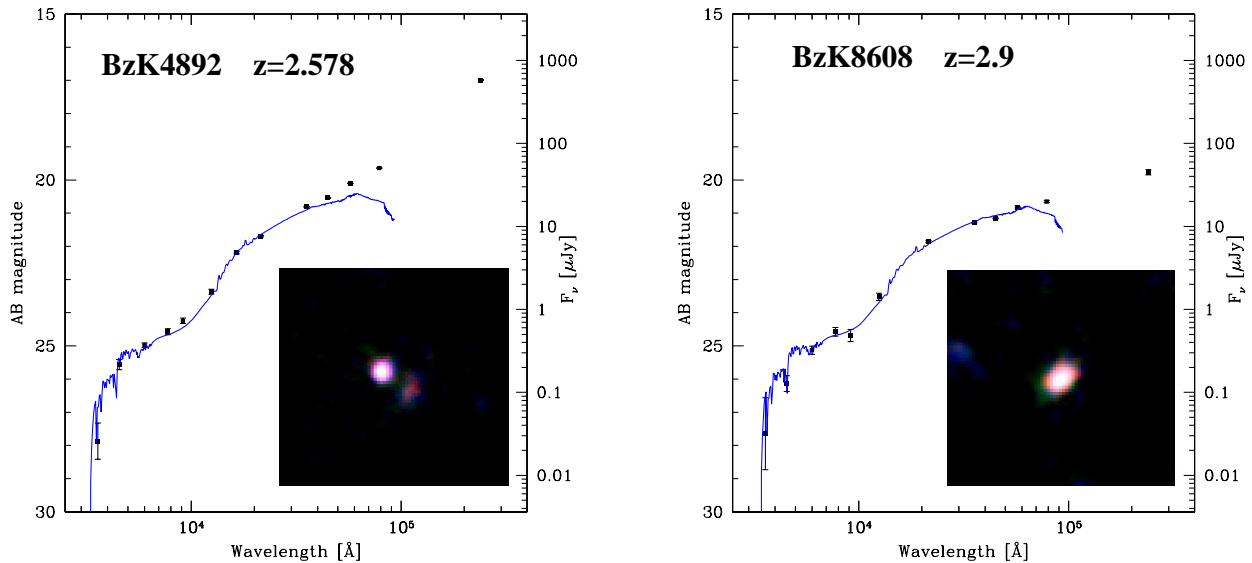}
\caption{Optical to mid-infrared SEDs of \bzk4892 (left panel, black symbols) and \bzk8608 (right panel). 
The blue
continuous line shows the fitted model SED of a pure star forming galaxy. The photometry from the GOODS datasets is taken from D07. HST+ACS color images are also shown as thumbnails. 
}
\label{fig:SED}
\end{figure*}

Both of the \bzk\ galaxies under exam were selected from the D07 sample. 
\bzk4892 is a source with spectroscopic redshift z$=$2.578 (Szokoly et al. 2004, Vanzella et al. 2006; 
based on multiple narrow emission lines), 
detected in the X-rays already in the 1 Ms \emph{Chandra} data (Giacconi et al. 2001) as well as in the 2 Ms data (Luo et al. 2010). 
In the 4 Ms dataset it has 75 net counts in the 0.5-7 keV band and hardness ratio of 0.58 (the hardness ratio is defined as HR$=$hard-soft/hard+soft, where S and H are the soft and hard band net counts detected by \emph{Chandra}).
\bzk8608  has a photometric redshift  z$_{phot}=2.94$ (from D07; consistent with  z$_{phot}=2.88$ from Santini
et al. 2009). 
VLT+VIMOS spectroscopic observations (Popesso et al. 2009) show that there is no continuum emission in the blue, consistent with z$\sim3$, but failed to yield a reliable redshift estimate.  
We tentatively detected in the VIMOS spectrum a low signal to noise feature at 6027 \AA\ (a region clear from OH lines) that, if identified with CIV, would give z$=2.88$.
We notice that the photometric redshift for \bzk8608 is reliable as the galaxy fulfills the U-band dropout criteria,
corresponding to a Lyman continuum break (being undetected in the deep U-band imaging of Nonino et al. 2010; 
Fig.~\ref{fig:SED}). 
This source is not listed in the \emph{Chandra} 2 Ms catalog (Luo et al. 2010), likely because of the presence 
of another nearby X-ray source (\#193).
In the 4 Ms data BzK8608 has 92 counts in the total band. It is not detected the 0.3-1.5 band, and its hardness ratio is
HR$>$0.54 (3$\sigma$). 

We extracted the \emph{Chandra} spectra using a circular region of 1.5$''$ and 3$''$ 
radius for BzK4892 and BzK8608 (off-axis $\sim$7.4\arcmin), respectively,
estimating the background in an annulus around the source position. We experimented with different extraction radii, 
centering, and background estimates, taking care in all cases of excluding nearby detected sources (particularly
important for BzK8608), finding that the
features found in the spectra were robust to the detail extraction method.
The spectra for both galaxies show a 
flat continuum and prominent emission lines peaking at $\sim1.78$~keV for \bzk4892 and $\sim1.64$~keV
for \bzk8608 (Fig. 1). 

We consider two spectral models to fit the data in XSPEC: i) a reflection-dominated model, 
\emph{pexrav + zgauss}, and ii) a transmission (power-law) model, \emph{wabs*(pow+zgauss)}.
A narrow gaussian component is used to model the emission lines.
Given the limited photon statistic, the slope of the primary continuum has been fixed to $\Gamma=1.8$ in both scenarios.
Based on C-statistics, the emission line is detected with a significance of  $\gtrsim$4$\sigma$ ($\Delta C=17.5$)  for \bzk4892  and $\gtrsim3\sigma$ ($\Delta C=7.5$) for \bzk8608 (see also Fig.~\ref{fig:Xzp}) for the reflection-dominated model. In the case of a transmission model the significance of the line remains above 4$\sigma$ for \bzk4892 and just below $3\sigma$ for \bzk8608.
To further evaluate the significance of the observed lines we also performed $10^5$ simulations of the source
continuum and background for both sources. The probability that the lines might be due to chance fluctuations 
are 3$\times10^{-5}$ and 2$\times10^{-4}$, respectively. The spectral fits are summarized in Table~1. 

Based on the spectroscopic and photometric redshifts, we identify the lines as Iron K$\alpha$ for both sources. 
In particular, we derive a spectroscopic redshift for \bzk8608 from the X-ray spectrum, obtaining $z=2.88\pm0.10$. 
Figure 2 shows the 1-2-3 $\sigma$ confidence level contours, showing the redshift fit and the significance of the Iron line detection.

The rest frame equivalent widths of the Iron lines are EW$=2.3_{-0.6}^{+1.0}$ keV for \bzk4892 and EW$=1.2\pm0.4$ keV for
\bzk8608 (for a reflection model, see Table~1), implying absorbing columns exceeding $10^{24}$ cm$^{-2}$.
\bzk8608 shows an additional low-significance excess at 1.8 keV that 
would correspond to the energy of Iron K$\beta$ at $z=2.9$, if real.
The detection of Iron K$\alpha$ at 6.4 keV excludes that the 
high hardness ratios in these sources might be due to high mass X-ray
binaries, as suggested by Donley et al. (2008), due to spectral incompatibility (Persic \& Raphaeli 2002).

The fit with the transmission model gives N$_H=7\pm2\times 10^{23}$ cm$^{-2}$ for \bzk4892 and $7^{+3}_{-2}\times
10^{23}$ cm$^{-2}$ for \bzk8608, and is statistically indistinguishable from the fit with a reflection model.
However, these N$_H$ values are inconsistent with the measured EW$\gtrsim$1 keV. 
Indeed, for N$_H\sim6\times 10^{23}$ cm$^{-2}$, a EW$\sim$300 eV is predicted (Ikeda et al. 2009) and confirmed by observations (Guainazzi et al.~2005, Fukazawa et al.~2010).  
EW$\gtrsim1$ keV are the hallmark of column densities larger than 10$^{24}$ cm$^{--}$.
Therefore, in the following, we adopt as best fit the results from the reflection dominated model for both sources
(Tab.~\ref{tab:data}).

Figure 2 shows the \emph{Chandra} images of \bzk4892 and \bzk8608  in three energy bands, including one centered
on the Iron lines.  We note that \bzk8608 is not detected in the soft band.

\begin{table*}
\begin{center}
\caption{Spectral fit parameters}
{\small
\begin{tabular}{lccccc}
\hline
Source ID & Model & Fe~K$\alpha$ flux & N$_H$ & EW  & Cstat/d.o.f.\\
    &    &  10$^{-17}$ erg/cm$^2$s  & 10$^{23}$ cm$^{-2}$ & keV  & \\
\hline
BzK4892 & Abs & 1.6$\pm$0.5  & 7.0$_{-2.0}^{+2.0}$   & 2.1$\pm$0.5 &  307/441\\
BzK4892 & Refl &   2.0$\pm$0.6 &  - & 2.3$_{-0.6}^{+1.0}$  &  297/442 \\
\hline
\hline
BzK8608 & Abs & 3.6$\pm$0.5  & 7.0$^{+3.0}_{-2.0}$ & 1.0$\pm$0.3 &   411/440\\
BzK8608 & Refl &  3.4$\pm$0.8 & - & 1.2$\pm$0.4 &  418/442 \\
\hline
 \end{tabular}
 }
\label{tab:fit}
\end{center}
\end{table*}

\section{Discussion}

\subsection{Intrinsic luminosities and obscuration}

The observed (obscured) hard X-ray luminosities are $L_{\rm 2-10~ keV}=10^{42.5}$ and
$10^{42.8}$~erg~s$^{-1}$ for \bzk4892 and \bzk8608, respectively. 
Given the large $N_{\rm H}$  implied by the detection
of high EW Iron K$\alpha$ lines, the X-ray emission is optically thick and we cannot directly estimate
the intrinsic $L_{\rm 2-10~ keV}$ using the X-ray data (e.g., La Massa et al. 2011).
In order to estimate the intrinsic luminosities of the AGN we used a variety of
independent methods. 

Following Iwasawa et al. (2005), a lower limit to the X-ray luminosity can be obtained in the expectation that 
in typical conditions the luminosity of the Iron K$\alpha$ line is at most 3\% of the 2-10 keV luminosity. 
For \bzk4892 L$_{K\alpha}=1.25\times10^{42}$~erg s$^{-1}$ would imply a intrinsic 2-10 keV luminosity of 4$\times10^{43}$~erg~s$^{-1}$.
The relation L$_{K\alpha}$/L$_{\rm 2-10~keV}$ of Levenson et al. (2006) implies an higher
luminosity, L$_{2-10~keV}\sim10^{44.8}$~erg~s$^{-1}$.
An alternative estimate is obtained from the 6.0\micron~ luminosity.   
The 24\micron~ flux of 591$\mu$Jy implies L$_{6.0\micron}=10^{45.5}$~erg~s$^{-1}$(the mid-IR SED of Mrk~231 is used for the small K-correction). While 24$\mu$m at $z=2.578$
might be contaminated by PAH features, we find similar results if we use the 16$\mu$m
flux of 209$\mu$Jy (Teplitz et al. 2010) for which no contamination is expected (4.5$\mu$m rest-frame).
This would suggest  L$_{\rm 2-10~ keV}=10^{45\pm0.5}$~erg~s$^{-1}$ using the Lutz et al. (2004) correlation and a lower L$_{\rm 2-10~
keV}=10^{44.4\pm0.5}$~erg~s$^{-1}$ using Lanzuisi et al. (2009).
A third estimate can be derived from the luminosity of the UV-optical emission lines.
The VLT/FORS1-2 optical spectrum of \bzk4892 shows prominent Ly$\alpha$, CIV, HeII and CIII] narrow emission lines (Szokoly et al. 2004, Vanzella et al. 2006).
Using the observed ratios between the line and the intrinsic 2-10~keV emission from Mulchaey et al. (1994),
we infer $L_{\rm 2-10~keV} = 10^{44.5\pm0.5}$~erg~s$^{-1}$.
Using the Netzer et al. (2006) conversion for UV-optical line luminosities would give larger X-ray 
luminosities by $\sim$0.5 dex, or L$_{\rm 2-10~ keV} = 10^{45\pm0.5}$~erg~s$^{-1}$.
All in all, it appears that the 2-10~keV luminosity of \bzk4892 is in the QSO range, well in excess of
$10^{44}$~erg~s$^{-1}$, and more likely of $10^{44.5-45}$~erg~s$^{-1}$. 
The implied column density is at the level of $N_{\rm H}\sim10^{24.5-25}$~cm$^{-2}$.

For \bzk8608, we measure L(K$\alpha$)$=2.5\times 10^{42}$~erg~s$^{-1}$. 
This converts to L$_{\rm 2-10~ keV}\sim10^{43.9}$~erg~s$^{-1}$ (Iwasawa et al. 2005), or $\sim10^{45}$~erg~s$^{-1}$ adopting Levenson et al. (2006). 
This galaxy is about 10 times fainter in the mid-IR than \bzk4892, which implies a lower L$_{6.0~\mu m}$. Using
the Lutz et al. (2004) relation, this in turn suggests L$_{\rm 2-10~ keV}=10^{44}$~erg~s$^{-1}$. We cannot obtain
an independent estimate from the UV-optical emission lines given that no lines were detected in the VIMOS
spectrum. 
The  intrinsic 2-10~keV luminosity of \bzk8608 is likely $\approx10^{44}$~erg~s$^{-1}$. 
The implied column density is likely in the range $N_{\rm H}\sim 1$--$5\times10^{24}$~cm$^{-2}$, hence this object
might be a {\em borderline} CT AGN.
We notice that both sources have estimated intrinsic 2-10 keV luminosities in the QSO regime.

\vspace{0.5cm}

\begin{table}[b]
\caption{Galaxy properties}
{\small
\begin{tabular}{lcc}
\hline
                                                                                                             &  \bzk4892                    &\bzk8608 \\ 
                                                                                                             \hline
 Ra                                                                                                      &  3:32:35.71                  &  3:32:20.95  \\
 Dec                                                                                                    & -27:49:16.1                 & -27:55:46.3 \\ 
 Redshift                                                                                                          & 2.578
 & 2.90$\pm0.10$   \\
 HR                                                                                                     &  0.58
 &  $>$0.54 ($3\sigma$) \\
 SFR$_{\rm UV}$ [M$_{\odot}$ yr$^{-1}]$               &  70                                 &  70 \\
 SFR$_{\rm Radio}$[M$_{\odot}$ yr$^{-1}]$              & 1100                                 &  300 \\
 L(K$\alpha$)    [erg~s$^{-1}$]                                           &  10$^{42.1}$                &   10$^{42.4}$ \\
 L$_{\rm 2-10~keV,~abs}$   [erg~s$^{-1}$]                   &  10$^{42.5}$                &   10$^{42.8}$\\
 L$_{\rm 2-10~keV,~K\alpha}$   [erg~s$^{-1}$]         &  10$^{43.6-44.8} $         &   10$^{43.9-45}$ \\
 L$_{\rm 2-10~keV,~6 \mu m}$ [erg~s$^{-1}$]         &   10$^{44.4-45}$         &   10$^{44}$ \\
 L$_{\rm 2-10~keV,~UV~ lines}$ [erg~s$^{-1}$]         &   10$^{44-45.5}$         &   - \\
 i [AB mag] & 24.75 & 24.72 \\
IRAC 3.6 $\mu$m [AB mag]  &  20.81   & 21.29\\
IRAC 4.5 $\mu$m [AB mag]  & 20.54    & 21.16 \\
IRAC 5.8 $\mu$m [AB mag]  & 20.1       & 20.84 \\
IRAC 8.0 $\mu$m [AB mag]  & 19.64    &  20.65 \\
MIPS 24 \micron~ [mJy] &  0.591$\pm0.07$  &   0.045$\pm0.003$\\
MIPS 70 \micron~ [mJy] &  3.3$\pm1.8$ &  -  \\      
 850 \micron~ [mJy]&   3.3$\pm$1.1     &   - \\
 1.4 GHz [$\mu$Jy] & 80$\pm$14 & 17.5$\pm6.5$\\
\hline
\hline
 \end{tabular}
}
\label{tab:data}
\end{table}

\subsection{Bolometric luminosities and star formation rates}

The ratio of the bolometric to the 2-10 keV luminosity,  L$_{bol}$/L$_{\rm 2-10~ keV}$, 
is typically of the order 30-50 for quasars (Marconi et al. 2004). 
Assuming such a bolometric correction,  we estimate 
L$_{bol}\sim 10^{46\pm0.5}$~erg~s$^{-1}$ for \bzk4892  and $\sim10^{45.5\pm0.5}$~erg~s$^{-1}$ for \bzk8608.
These very high luminosities, if expressed in solar units, correspond to L$_{bol}\sim10^{12.4\pm0.5}L_\odot$
and $\sim10^{11.9\pm0.5}L_\odot$.  When powered by star formation,
similar luminosities require star formation rates (SFR) in the host galaxies at the level of 
100-1000~M$_\odot$~yr$^{-1}$. 
It is interesting thus to compare these estimates to the inferred SFR for the two \bzk\ galaxies, in order to
compare the relative AGN and SFR contributions.

From the stellar SED (UV to the near-IR rest frame; Fig.~3), 
we estimate $SFR\sim70$~M$_\odot$~yr$^{-1}$
for \bzk4892, corrected for dust reddening based on the observed UV slope (see D07 for more details). 
It is quite possible though that also the stellar UV emission
might be optically thick.
If we interpret its mid-IR emission (591 $\mu$Jy at 24$\mu$m) as due to star
formation this would imply a whopping SFR$\sim10^4$~M$_\odot$~yr$^{-1}$, demonstrating that the 
mid-IR is likely completely dominated by the AGN (this object is 
among the most extreme mid-IR excess galaxies in the sample of D07). 
The SED fit with a star forming galaxy template (Fig.~3)
shows an excess emission already at 5.8 \micron~ (observed frame) probably due to the AGN contribution.  
The SED in the IRAC bands is indeed showing a steep power law, extending all the way to 24 \micron.
Due to its faint optical counterpart (i$\sim$25 AB), \bzk4892 has high mid-infrared to optical flux ratio, F(24)/F(R)$\sim$2000 and therefore is also classified as a dust-obscured galaxy (DOG, Dey et al. 2008). 
It is an extreme object also in the Fiore et al. (2008) sample.  
Alonso-Herrero et al. (2006) reports a
70 \micron~  flux of $3.3\pm1.8$ mJy for this galaxy, which is also likely affected by the AGN emission. 
On the other hand, the galaxy is seen at 1.4 GHz with a flux of 80~$\mu$Jy in the VLA data of Miller et al. (2008). 
If due to star formation,
using the radio-IR correlation,
this would imply SFR$\sim1100$~M$_\odot$~yr$^{-1}$. Inspecting the publicly available Apex+LABOCA 870\micron~ map of GOODS-S
(Weiss et al. 2009), we find a 3$\sigma$ signal at the position of \bzk4892 of 3.3~mJy, which would also convert into a
similar SFR, $\sim500$~M$_\odot$~yr$^{-1}$. 

Even for the most luminous quasars it is generally found that the
far-IR emission in the sub-mm bands is due to star formation.  If this is the case also for \bzk4892, our
results suggest that the galaxy is witnessing also very powerful star formation activity at the level of
500-1000~M$_\odot$~yr$^{-1}$. Only 5-10\% of this is seen directly in the UV, implying that also the UV-emission
from stars is optically thick, similarly to local ULIRGs (e.g., Goldader et al. 2002; da Cunha et al. 2010)
and to what is expected to be found in major mergers.
However, we notice that the radio and 870$\mu$m derived SFRs would formally imply
L$_{bol}\sim 10^{46-46.5}$~erg~s$^{-1}$, comparable to what inferred for the obscured AGN. 
Hence, the AGN is likely contributing an important fraction of the total L$_{bol}$.

The morphology of the galaxy from HST+ACS imaging is suggestive of a merger, showing two distinct clumps in the
UV in the rest frame. When smoothing, after excluding the two bright knots, we detect significant faint low-level
emission in the summed $i+z$ band, extended over a
diameter of about 0.75\arcsec (6~kpc). Overall, this is consistent with the size of a  massive galaxy at 
z$\sim 2$, and we cannot exclude that the two UV knots might be just luminous HII regions inside
a big disk galaxy.

The photometric information is of lower quality for \bzk8608 due to its overall faintness. 
The SFR inferred from the UV is also 70~M$_{\odot}/yr$.
This object is also a mid-IR excess source in D07 sample (by a less extreme factor of 6), 
with S$_{24} = 45 ~\mu$Jy ( $\sim$1/10 of \bzk4892), although it is not a DOG.
The SED (Fig.~3) 
might also be consistent with pure stellar emission, but a deviation from the star formation template 
is observed at the $8~\mu$m band.
This galaxy
is not detected at 70 \micron~ and 870 \micron, suggesting a lower AGN luminosity and/or SFR. Inspecting 
the VLA 1.4 GHz data (Miller et al. 2008) we detect a faint $3\sigma$ source at its position with
a flux of $17.5\pm6.5~\mu$Jy. This corresponds to a luminosity $3\times10^{12}$ L$_{\odot}$ and
SFR$\sim300$~M$_\odot$~yr$^{-1}$. If the radio signal is real and not due to AGN, also in this case the stellar
emission appears to be heavily obscured, i.e. optically thick in the UV, as expected for very active
starbursts and mergers. The HST+ACS imaging of this galaxy (Fig.~3) is not
particularly telling.

For both galaxies we derive similar estimates of the stellar masses, at the level of $5\times10^{10}M_\odot$. 
Our results thus support the picture that the luminous, Compton thick AGN that we discovered are hosted by
fairly massive galaxies at $z\sim2.5$-3, that at the same time also host vigorous star formation activity
heavily obscured by dust.
This is relevant in the AGN-host galaxy co-evolution scenario, in which a phase of rapid, heavily obscured BH growth accompanied by intense 
obscured star-formation is predicted (Silk \& Rees 1998, Fabian 1999, Granato et al. 2004).

\subsection{Implications for Compton thick nuclear activity at high redshift}

We derive a crude value of the volume density of Compton thick AGN with high EW
Iron K$\alpha$ emission, based on our 2 detections. We 
conservatively use the full redshift range explored by the \bzk~ selection 
(z$=$1.4-3),  finding a space density of $3\times 10^{-6}$ Mpc$^{-3}$. This is  
consistent with the predictions of the Gilli et al. (2007) X-ray background synthesis model, for Compton thick
AGN with L$_{2-10~ keV}>10^{44}$~erg~s$^{-1}$. However, we notice that our sampling of L$_{2-10~keV}>10^{44}$~erg~s$^{-1}$ 
AGN with Compton thick absorption at $1.4<z<3$ might be substantially incomplete 
due to  selection effects, and the luminosity of at least one of the 
\bzk~ galaxies in this paper might be substantially higher
than the adopted $10^{44}$~erg~s$^{-1}$ limit.

We emphasize that the sources presented here have X-ray spectra that are the high-luminosity analogous of those
of local prototype Compton thick AGN. This discovery represents one of the first 
clear-cut evidence for this class of objects at high redshift.
Our results confirm that indeed there is a population of Compton thick QSO among massive, star-forming, dust obscured galaxies, as suggested by several studies (D07, Fiore et al. 2008, 2009, Treister et al. 2010, Alexander et al. 2008, Lanzuisi et al. 2009), and as predicted by models of merger-driven AGN/galaxy co-evolution (Silk \& Rees 1998, Fabian 1999, Di Matteo et al. 2005). 
Indeed, these luminous  Compton thick quasars appear to be hosted
by galaxies with optically thick UV emission, which are a relative small
minority among BzKs (Daddi et al 2007a), and imply the presence of a
dense starburst usually connected with mergers.
This seems to agree with evolutionary scenarios, where mergers induce both rapid accretion onto SMBH and intense star-formation activity.

The two sources presented here have high intrinsic X-ray and bolometric  luminosity, and bright Iron K$\alpha$ line,
allowing their identification. They might represent just the tip of the iceberg of a much larger
population of highly obscured AGN with somewhat lower intrinsic luminosity, whose Iron lines are still remaining
hardly detectable even in the deepest hard X-ray surveys.  For example, the Gilli et al. model predicts a
$\sim10\times$ higher space density of Compton thick sources with L$_{\rm 2-10~ keV}>10^{43}$~erg~s$^{-1}$. 
Detecting Iron lines in galaxies with such luminosities might be challenging with current facilities,
even if they have similar EW as in \bzk4892 and \bzk8608.

\acknowledgments
We thank the anonymous referee for useful comments.
This research was
supported by the ERC-StG grant UPGAL 240039 and by the
French ANR under contract ANR-08-JCJC-0008.
EP acknowledges financial support from ASI (grant I/088/06/0). 
DMA thanks the Science 
and Technology Research Council for support.

\end{document}